\documentclass[sigplan,10pt,nonacm]{acmart}

\usepackage{amsmath}
\usepackage{array}
\usepackage{booktabs}
\usepackage{enumitem}
\usepackage{float}
\usepackage{graphicx}
\usepackage{listings}
\usepackage{needspace}
\usepackage{placeins}
\usepackage{tikz}
\usepackage{xcolor}
\graphicspath{{figures/}}

\renewcommand\footnotetextcopyrightpermission[1]{}
\setlist{nosep,leftmargin=*}
\newcommand{\tool}{KernelScript}
\newcommand{\code}[1]{\texttt{\detokenize{#1}}}

\title{\tool: Cross-Boundary Typed DSL for eBPF Applications}
\author{Cong Wang$^{1}$, Siyuan Sun$^{2}$, Yusheng Zheng$^{3}$}
\affiliation{\institution{$^{1}$Multikernel Technologies \quad $^{2}$Xi'an University of Posts and Telecommunications \quad $^{3}$UC Santa Cruz \& Eunomia Labs}\country{}}
\email{cwang@multikernel.io, a879650736@gmail.com, yzhen165@ucsc.edu}
\renewcommand{\authorsaddresses}{}
\renewcommand{\shortauthors}{Wang et al.}
\begin{abstract}
eBPF lets developers extend Linux with custom packet processing, tracing, and scheduling logic, and a verifier proves before execution that the code will not crash the kernel.
The programming model, however, is fragmented: a single application spans kernel code, a userspace loader, and shared maps, yet the relationships among these pieces go unchecked.
E.g. A map or event type defined differently on each side silently corrupts shared state.
We observe that these cross-boundary relationships duplicate information that a type system can unify.
We present \tool{}, a DSL that types maps, program handles, and execution domains in one source, then compiles to standard C through the original toolchain.
We evaluate \tool{} on 43 eBPF workloads covering XDP, TC, kprobe, tracepoint, and struct\_ops.
\tool{} rejects cross-boundary bugs at compile time that standard C/libbpf still builds and loads, a unified source shrinks the diffs for cross-boundary changes by 5$\times$, and generated code remains compatible with the existing toolchain.
\end{abstract}


\begin{document}
\maketitle
\pagestyle{plain}

\section{Introduction}
% 引言

% P1: Context/Importance
eBPF lets developers run application-specific logic inside the Linux kernel under a verifier that proves memory safety and bounded execution~\cite{linux-ebpf}.
% eBPF 让开发者在 Linux 内核内运行应用程序特定的逻辑，由验证器证明内存安全和有界执行。
It now underpins packet processing in major cloud providers, powers observability platforms, and enables scheduler experimentation, all within a large and growing ecosystem~\cite{ebpfsurvey}.
% 它现在支撑着主要云提供商的数据包处理，驱动可观测性平台，并在一个庞大且不断增长的生态系统中支持调度器实验。

% P2: Problem
Yet the programming model is fragmented.
% 然而编程模型是碎片化的。
A single eBPF application spans kernel code constrained by the verifier, a userspace loader that opens objects and manages lifecycle, shared maps and ring buffers, and generated headers that must match the running kernel.
% 单个 eBPF 应用程序跨越受验证器约束的内核代码、打开对象并管理生命周期的用户空间加载器、共享映射和环形缓冲区，以及（对于较新功能）必须与运行中的内核匹配的生成头文件。
The relationships among these pieces, such as a map's key and value types or the order of load, attach, and detach, are checked only at load time if at all.
% 这些部分之间的关系，例如映射的键和值类型或加载、附加和分离的顺序，仅在加载时检查或根本不检查。
Errors may go entirely undetected because they do not break the kernel.
% 错误可能完全不被检测，因为它们不会破坏内核。
When a map's value struct is redefined in kernel C but not in the userspace loader, data is silently misinterpreted at runtime.
% 当映射的值结构体在内核 C 中被重新定义但用户空间加载器未更新时，数据在运行时被静默误解。
When a generated header becomes stale after a struct change, userspace compiles against the old layout and reads corrupt fields.
% 当生成头文件在结构体更改后变得过时，用户空间针对旧布局编译并读取损坏的字段。

% P3: Gap
Existing tools reduce boilerplate yet leave cross-boundary relationships implicit.
% 现有工具减少了样板代码，但跨边界关系仍然是隐式的。
The standard eBPF development workflow requires hand-written kernel C, userspace C, and a Makefile~\cite{libbpf}.
% 标准的 eBPF 开发工作流需要手写内核 C、用户空间 C 和 Makefile。
Aya compiles both sides from Rust and shares typed maps~\cite{aya}, and cilium/ebpf provides the same in Go~\cite{cilium-ebpf}.
% Aya 从 Rust 编译两端并共享类型化映射，cilium/ebpf 在 Go 中做类似的事情。
In both, map types are shared but other cross-boundary invariants remain implicit: code that attaches before populating a map compiles without error.
% 在两者中，映射类型是共享的，但其他跨边界不变量仍然是隐式的：在填充映射之前附加编译时不报错。
No existing tool types the full cross-boundary structure at compile time.
% 没有现有工具在编译时对完整的跨边界结构进行类型化。

% P4: Insight
The key observation is that these relationships are single concepts duplicated across components.
% 关键观察是这些关系是跨组件重复的单一概念。
A map declaration is simultaneously a kernel data structure, a userspace lookup target, and a type contract.
% 映射声明同时是内核数据结构、用户空间查找目标和类型契约。
A program handle is simultaneously a function name, a BPF section, and a file descriptor.
% 程序句柄同时是函数名、BPF 节和文件描述符。
In C/libbpf, naming conventions connect these copies, so mismatches go undetected until load time or runtime.
% 在 C/libbpf 中，命名约定连接这些副本，因此不匹配延迟出现。
In a unified typed language, the compiler can see and check these relationships directly.
% 在统一的类型化源文件中，编译器可以直接看到并检查它们。

% P5: Solution
\tool{} implements this idea as a compiled DSL that spans kernel and userspace.
% \tool{} 将这个想法实现为一个跨越内核与用户空间的编译型 DSL。
A developer writes eBPF entry points, userspace control flow, maps, and ring buffers in one source file.
% 开发者在一个源文件中编写 eBPF 入口点、用户空间控制流、映射和环形缓冲区。
The compiler types maps, program handles, and execution domains, then emits eBPF C, userspace C, and a Makefile.
% 编译器对映射、程序句柄和执行域进行类型化，然后生成 eBPF C、用户空间 C 和 Makefile。
The standard libbpf and bpftool path then produces the final binaries, preserving the existing toolchain.
% 标准的 libbpf 和 bpftool 路径然后生成最终二进制文件，保留现有工具链。

% P6: Contributions
This paper contributes:
% 本文贡献：
\begin{itemize}
\item A typed model of cross-boundary eBPF structure: execution domains, typed maps, and program handles with lifecycle rules, each with a typing rule and matching compiler diagnostic (Section~\ref{sec:model}).
% 跨边界 eBPF 结构的类型化模型：执行域、类型化映射和带生命周期规则的程序句柄，每个都有类型规则和匹配的编译器诊断（第~\ref{sec:model} 节）。
\item A compiler that implements this model while preserving the existing kernel toolchain (Section~\ref{sec:impl}).
% 实现此模型同时保留现有内核工具链的编译器（第~\ref{sec:impl} 节）。
\item Evaluation showing that \tool{} rejects bugs at compile time that C/libbpf still builds and loads, produces smaller diffs for cross-boundary changes, and remains compatible with clang, bpftool, and libbpf (Section~\ref{sec:eval}).
% 评估包括：对 C/libbpf 仍能构建并加载的 bug 的编译期拒绝、相对 C/libbpf 同一修改更小的 diff，以及在真实内核上与 clang、bpftool 和 libbpf 的兼容性（第~\ref{sec:eval} 节）。
\end{itemize}

\section{Background and Motivation}
% 背景与动机
\label{sec:bg}

eBPF allows user-supplied bytecode to run inside the Linux kernel after a verifier proves memory safety and bounded execution~\cite{linux-ebpf,ebpfsurvey}.
% eBPF 允许用户提供的字节码在验证器证明内存安全和有界执行后在 Linux 内核内运行。
The mechanism now underpins a wide range of applications, including tracing and profiling~\cite{bpftrace,bcc}, high-performance networking~\cite{xdp}, LSM policy enforcement~\cite{lsm-bpf}, CPU scheduling~\cite{ghost}, page-cache customization~\cite{cacheext}, GPU resource control~\cite{gpuext}, and userspace extension runtimes~\cite{bpftime}.
% 该机制现在支撑着广泛的应用：追踪和性能分析、高性能网络、LSM 策略执行、CPU 调度、页缓存自定义、GPU 资源控制和用户空间扩展运行时。
Every eBPF application has a kernel program and a userspace loader that communicate through maps or ring buffers.
% 每个 eBPF 应用程序都有一个内核程序和一个通过映射或环形缓冲区通信的用户空间加载器。
The standard eBPF workflow requires maintaining separate kernel C, userspace C, and a Makefile~\cite{libbpf}.
% 标准的 eBPF 开发工作流要求开发者维护单独的内核 C、用户空间 C 和 Makefile。

The kernel/userspace split introduces problems that current tools do not fully address.
% 内核/用户空间分离引入了当前工具未完全解决的问题。
\emph{Lifecycle ordering}: operations such as load, attach, and detach must occur in a valid order, but the raw libbpf API checks this only at runtime.
% 生命周期顺序：诸如 load、attach 和 detach 的操作必须以有效顺序发生，但 C/libbpf 仅在运行时检查。
\emph{Shared types}: a map's key and value types must agree across kernel code, generated headers, and userspace lookups, yet a mismatch is silent until something fails.
% 共享类型：映射的键和值类型必须在内核代码、生成的头文件和用户空间查找之间一致，然而不匹配在出现故障之前是静默的。
\emph{Version mismatches} arise because generated bindings depend on the exact kernel and libbpf versions.
% 版本不匹配出现是因为生成的绑定依赖确切的内核和 libbpf 版本。
The verifier guarantees memory safety but not application correctness across the boundary.
% 验证器保证内存安全，但不保证跨边界的应用正确性。
Calling \code{attach} before populating maps lets the kernel program run against empty state.
% 在填充映射之前调用 \code{attach} 会使内核程序在空状态下运行。
Redefining a ring-buffer event struct on one side without updating the other silently corrupts every record.
% 在一侧重新定义环形缓冲区事件结构体而不更新另一侧会静默破坏每条记录。
A stale generated header after a kernel struct change compiles without warning but misaligns every shared field.
% 内核结构体更改后过时的生成头文件在没有警告的情况下编译，但使每个共享字段对齐错误。
A DSL can check load/attach order and shared map types at compile time.
% DSL 可以在编译时检查 load/attach 顺序与共享映射类型。

\section{A Typed Model of Cross-Boundary Structure}
% 跨边界结构的类型化模型
\label{sec:model}

The type system draws on established ideas.
% 类型系统借鉴了已建立的思想。
Typestate tracks object state in types~\cite{typestate}, and linear types enforce single-use resource protocols~\cite{lineartypes,vault,sessiontypes}.
% 类型状态在类型中跟踪对象状态，线性类型强制执行单次使用资源协议。
\tool{} applies a lightweight form of these ideas, separating unloaded function references from loaded program handles and annotating execution domains without full linear or session types.
% \tool{} 应用这些思想的轻量级形式，区分未加载的函数引用与已加载的程序句柄并标注执行域，而不做完整的线性或会话类型。

A \tool{} program is a sequence of declarations.
% \tool{} 程序是声明的序列。
Listing~\ref{lst:min} shows a minimal application.
% 清单~\ref{lst:min} 展示一个最小应用程序。
The rest of this section describes domains, maps, and handles.
% 本节其余部分说明域、映射和句柄。

% 统一源 eBPF 应用程序示例。
\begin{lstlisting}[language=C,caption={A unified-source eBPF application.},label={lst:min}]
include "xdp.kh"

@xdp fn pass_all(ctx: *xdp_md) -> xdp_action {
  return XDP_PASS
}

fn main() -> i32 {
  var prog = load(pass_all)
  attach(prog, "lo", 0)
  detach(prog)
  return 0
}
\end{lstlisting}

To check execution domains, each function carries an attribute that assigns it to an execution domain $d \in \{\textit{user}, \textit{ebpf}, \textit{helper}, \textit{kfunc}\}$.
% 为了检查执行域，每个函数都带有一个属性，将其分配到执行域 $d \in \{\textit{user}, \textit{ebpf}, \textit{helper}, \textit{kfunc}\}$。
The attributes \code{@xdp}, \code{@tc}, \code{@probe}, \code{@tracepoint}, and \code{@perf_event} mark eBPF entry points, while \code{@helper} marks kernel-shared eBPF code and \code{@kfunc} marks kernel-module code.
% 属性 \code{@xdp}、\code{@tc}、\code{@probe}、\code{@tracepoint} 和 \code{@perf_event} 标记 eBPF 入口点，而 \code{@helper} 标记内核共享的 eBPF 代码，\code{@kfunc} 标记内核模块代码。
An unattributed \code{fn} is userspace.
% 没有属性的 \code{fn} 是用户空间。
The checker tags each function with its domain and rejects illegal crossings, such as a userspace function calling an \code{@helper}.
% 检查器用其域标记每个函数，并拒绝非法跨越，例如用户空间函数调用 \code{@helper}。
Each eBPF attribute also fixes a context and return type.
% 每个 eBPF 属性还固定上下文和返回类型。
For example, \code{@xdp} requires \code{ctx:*xdp_md} returning \code{xdp_action}, and \code{@tc} requires \code{*__sk_buff} returning \code{i32}.
% 例如，\code{@xdp} 需要 \code{ctx:*xdp_md} 返回 \code{xdp_action}，\code{@tc} 需要 \code{*__sk_buff} 返回 \code{i32}。
The compiler rejects any signature that violates its attribute before code generation.
% 编译器在代码生成之前拒绝任何与其属性不一致的签名。
Userspace can allocate and print freely, while eBPF code must obey stack and helper restrictions.
% 用户空间可以自由分配和打印，而 eBPF 代码必须遵守栈和辅助函数限制。
struct\_ops callbacks fit the same scheme, with context and return types determined by the registered slot.
% struct\_ops 回调适合相同的方案，上下文和返回类型由注册的槽位确定。

To keep map types consistent on both sides, a map is declared as a typed global (\code{var m : hash<K,V>(N)}) rather than a C section plus separate userspace lookup code.
% 为了使两侧映射类型一致，映射被声明为类型化全局变量（\code{var m : hash<K,V>(N)}），而不是 C 节加单独的用户空间查找代码。
The compiler checks element types at every access site on both sides of the kernel boundary. An index \code{m[k]} requires \code{k:K} and yields \code{V}, in eBPF and userspace alike.
% 编译器在内核边界两侧的每个访问点检查元素类型：索引 \code{m[k]} 需要 \code{k:K} 并产生 \code{V}，在 eBPF 和用户空间中都是如此。
A wrong key type, wrong value type, or access to an undeclared map is a compile-time error.
% 错误的键类型、错误的值类型或访问未声明的映射是编译时错误。
One declaration thus replaces the kernel definition, generated fields, and userspace lookups that would otherwise require manual synchronization.
% 因此，一个声明替换了内核定义、生成的字段和用户空间查找，否则这些必须手动保持一致。

To enforce load before attach, the type system distinguishes a \code{FunctionRef} (a reference to an \code{@}-attributed function) from a \code{ProgramHandle} (a loaded program).
% 为了强制先 load 再 attach，类型系统区分 \code{FunctionRef}（对 \code{@} 属性函数的引用）和 \code{ProgramHandle}（已加载的程序）。
Only \code{load} produces a \code{ProgramHandle}.
% 类型化确保只有 \code{load} 产生 \code{ProgramHandle}。
The \code{FunctionRef} type connects lifecycle to domains.
% \code{FunctionRef} 的类型规则将生命周期连接到域。
Only an eBPF entry-point function has type \code{FunctionRef}, so an \code{@helper}, \code{@kfunc}, or userspace function is rejected as an argument to \code{load}.
% 只有 eBPF 入口点函数具有类型 \code{FunctionRef}，因此 \code{@helper}、\code{@kfunc} 或用户空间函数作为 \code{load} 的参数会被拒绝。
\begin{gather*}
\frac{f \text{ has an eBPF entry attribute}}
     {\Gamma \vdash f : \texttt{FunctionRef}}
\\[3pt]
\frac{\Gamma \vdash f : \texttt{FunctionRef}}
     {\Gamma \vdash \texttt{load}(f) : \texttt{ProgramHandle}}
\\[3pt]
\frac{\Gamma \vdash h : \texttt{ProgramHandle}}
     {\Gamma \vdash \texttt{detach}(h) : \texttt{unit}}
\\[3pt]
\frac{\Gamma \vdash h : \texttt{ProgramHandle}}
     {\Gamma \vdash \texttt{attach}(h, t, \mathit{fl}) : \texttt{u32}}
\end{gather*}
Because \code{attach} and \code{detach} require a \code{ProgramHandle} and only \code{load} produces one, ``attach before load'' is a \emph{type} error, not a runtime failure.
% 因为 \code{attach} 和 \code{detach} 需要 \code{ProgramHandle} 而只有 \code{load} 产生它，"加载前附加"是一个类型错误，而不是运行时故障。
Passing a bare \code{FunctionRef}, an integer, or a string to \code{load}, \code{attach}, or \code{detach} is a type error.
% 将裸 \code{FunctionRef}、整数或字符串传给 \code{load}、\code{attach} 或 \code{detach} 是类型错误。
Because only \code{load} produces a \code{ProgramHandle}, the system enforces the single ordering constraint $\textit{load} \prec \{\textit{attach},\,\textit{detach}\}$ rather than the full load, attach, then detach sequence.
% 因为只有 \code{load} 产生 \code{ProgramHandle}，系统强制执行单一排序约束 $\textit{load} \prec \{\textit{attach},\,\textit{detach}\}$，而不是完整的 load、attach、再 detach 序列。
Full path-sensitive properties such as use-after-detach, double-attach, or leaked handles remain outside the type system's scope.
% 完整的路径敏感属性如分离后使用、双重附加或泄漏的句柄仍在类型系统范围之外。

\section{Design and Implementation}
% 设计与实现
\label{sec:impl}

The compiler has three design goals.
% 编译器有三个设计目标。
First, it must detect cross-boundary errors at compile time, before the verifier or runtime sees them.
% 首先，它必须在编译时检测跨边界错误，在验证器或运行时看到它们之前。
Second, it must preserve the existing Linux BPF toolchain.
% 其次，它必须保留现有的 Linux BPF 工具链。
Third, it must generate idiomatic C that developers can inspect and debug.
% 第三，它必须生成开发者可以检查和调试的惯用 C 代码。

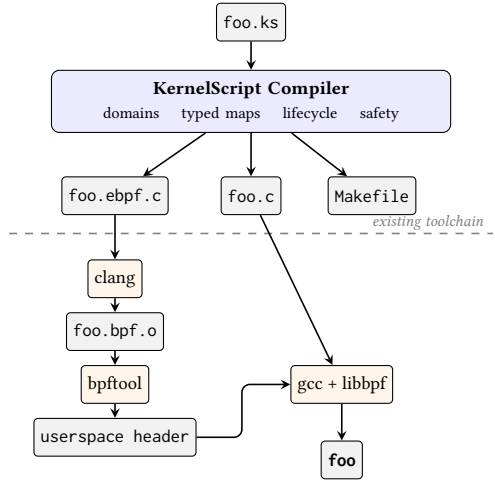
\begin{figure}[t]
\centering
\begin{tikzpicture}[
  font=\scriptsize,
  file/.style={draw, rounded corners=1.5pt, fill=gray!10,
    inner sep=2.5pt, font=\scriptsize\ttfamily, minimum height=5mm},
  phase/.style={draw, rounded corners=3pt, fill=blue!8,
    inner sep=4pt, align=center, text width=50mm},
  tool/.style={draw, rounded corners=1.5pt, fill=orange!8,
    inner sep=2.5pt, minimum height=5mm},
  arr/.style={->, >=stealth, semithick},
]

\node[file] (src) at (0, 0) {foo.ks};

\node[phase] (comp) at (0, -1.05) {%
  \textbf{\tool{} Compiler}\\[1pt]
  {\tiny domains \quad typed maps \quad lifecycle \quad safety}};
\draw[arr] (src) -- (comp);

\node[file] (ebpfc) at (-1.8, -2.3) {foo.ebpf.c};
\node[file] (userc) at (0, -2.3) {foo.c};
\node[file] (makef) at (1.6, -2.3) {Makefile};
\draw[arr] (comp) -- (ebpfc);
\draw[arr] (comp) -- (userc);
\draw[arr] (comp) -- (makef);

\draw[dashed, gray, semithick] (-3.2, -2.8) -- (3.2, -2.8);
\node[anchor=east, gray, font=\tiny\itshape] at (3.2, -2.65) {existing toolchain};

\node[tool] (clang) at (-1.8, -3.4) {clang};
\draw[arr] (ebpfc) -- (clang);

\node[file] (bpfo) at (-1.8, -4.1) {foo.bpf.o};
\draw[arr] (clang) -- (bpfo);

\node[tool] (bpftool) at (-1.8, -4.8) {bpftool};
\draw[arr] (bpfo) -- (bpftool);

\node[file] (skel) at (-1.8, -5.5) {userspace header};
\draw[arr] (bpftool) -- (skel);

\node[tool] (gcc) at (1.2, -4.8) {gcc + libbpf};
\draw[arr] (userc) -- (gcc);
\draw[arr, rounded corners=3pt] (skel.east) -- ++(0.6, 0) |- (gcc.west);

\node[file, font=\scriptsize\ttfamily\bfseries] (bin) at (1.2, -5.8) {foo};
\draw[arr] (gcc) -- (bin);

\end{tikzpicture}
\caption{Compilation pipeline. The compiler checks cross-boundary
relationships and emits C, and the existing clang/bpftool/libbpf toolchain
produces the final binary.}
% 编译流水线。编译器检查跨边界关系并发出 C，现有 clang/bpftool/libbpf 工具链生成最终二进制。
\label{fig:pipeline}
\end{figure}

To meet these goals, the compiler parses \tool{} source, resolves imports and includes, builds symbol tables, type-checks across programs, checks safety constraints, builds an IR, and emits C (Figure~\ref{fig:pipeline}). The compiler is written in OCaml with dune.
% 为了实现这些目标，编译器（用 OCaml 和 dune 编写）解析 \tool{} 源代码，解析导入和包含，构建符号表，跨程序做类型检查，检查安全约束，构建 IR，并发出 C 代码（图~\ref{fig:pipeline}）。
The compiler comprises nine modules covering maps, codegen, safety, and struct\_ops.
% 编译器包含九个模块，涵盖映射、代码生成、安全性和 struct\_ops。
For an input \texttt{foo.ks}, the generated project contains \texttt{foo.ebpf.c}, \texttt{foo.c}, a \texttt{Makefile}, and, when kfuncs are present, \texttt{foo.mod.c}.
% 对于输入 \texttt{foo.ks}，生成的项目包含 \texttt{foo.ebpf.c}、\texttt{foo.c}、\texttt{Makefile}，以及当存在 kfuncs 时的 \texttt{foo.mod.c}。
The Makefile invokes bpftool to dump \texttt{vmlinux.h}, clang to produce a BPF object, bpftool to generate a userspace header, and gcc to link the userspace loader against libbpf.
% Makefile 调用 bpftool 转储 \texttt{vmlinux.h}，clang 生成 BPF 对象，bpftool 生成用户空间头文件，gcc 将用户空间加载器与 libbpf（以及 libelf 和 zlib）链接。

Each typing rule of Section~\ref{sec:model} maps to a compiler pass.
% 第~\ref{sec:model} 节的每条类型规则映射到一个编译器遍。
The IR records each function's execution domain so that code generation can dispatch the same source function to the eBPF or userspace backend.
% IR 记录每个函数的执行域，以便代码生成将同一源函数分派到 eBPF 或用户空间后端。
Typed maps compile to a single descriptor that both backends read to emit the eBPF \code{SEC(".maps")} definition and the userspace accessor (Listing~\ref{lst:map}).
% 类型化映射编译为单个描述符，两个后端读取该描述符以发出 eBPF \code{SEC(".maps")} 定义和用户空间访问器（清单~\ref{lst:map}）。
Program handles become file descriptors via libbpf APIs at runtime.
% 程序句柄在运行时通过 libbpf API 变为文件描述符。

% 一个类型化映射声明从同一描述符同时生成内核 SEC(".maps") 定义与用户空间访问器。
\begin{lstlisting}[language=C,caption={One typed map declaration emits both
a kernel \texttt{SEC(".maps")} definition and a userspace accessor from a single
descriptor.},label={lst:map}]
// unified source
var counts : hash<u32, u64>(1024)

// generated eBPF C
struct {
  __uint(type, BPF_MAP_TYPE_HASH);
  __uint(max_entries, 1024);
  __type(key, __u32);
  __type(value, __u64);
} counts SEC(".maps");

// generated userspace C: same key/value types
__u32 key = ...; __u64 val;
bpf_map_lookup_elem(counts_fd, &key, &val);
\end{lstlisting}

The compiler does not replace the verifier, clang, bpftool, or libbpf, so their diagnostics remain available.
% 编译器不替换验证器、clang、bpftool 或 libbpf，因此其诊断仍可用。
For fast-moving interfaces, \code{include} and \code{extern} declarations provide a fallback to raw headers.
% 对于比 DSL 稳定更快移动的接口，\code{include} 和 \code{extern} 声明提供回退到原始头文件声明。

\section{Evaluation}
% 评估
\label{sec:eval}

All experiments run on Linux 6.15 with clang 18, bpftool 7.7, and libbpf 1.3.0.
% 所有实验在 Linux 6.15 上运行，使用 clang 18、bpftool 7.7 和 libbpf 1.3.0。

\subsection{Q1: Does \tool{} Reject Bugs at Compile Time that C/libbpf Still Loads?}
% Q1：\tool{} 是否在编译期拒绝 C/libbpf 仍能加载的 bug？

A rejection suite of 31 buggy programs (28 constructed, 3 from prior work~\cite{heimdall}) covers the typing rules of Section~\ref{sec:model}.
% 31 个错误程序的拒绝套件（28 个自构造，3 个来自 prior work）覆盖第~\ref{sec:model} 节的类型规则。
The 31 rejections break down as: 7 lifecycle/handle misuses (e.g., \code{attach} on \code{FunctionRef}, \code{load("string")}), 6 wrong context or return types (e.g., \code{@xdp} with \code{*__sk_buff}), 6 map type errors (e.g., string key, value size mismatch), 3 illegal domain crossings (e.g., userspace calling \code{@helper}), 4 general type errors (e.g., \code{42+true}), 3 safety violations (e.g., 600-byte stack), and 2 perf-event constraints.
% 31 个拒绝分解为：7 个生命周期/句柄误用，6 个错误的上下文或返回类型，6 个映射类型错误，3 个非法域跨越，4 个一般类型错误，3 个安全违规，2 个 perf-event 约束。
All fail before code generation with a diagnostic message matching the expected error.
% 全部在代码生成前以匹配预期错误的诊断消息失败。

For the three prior-work cases, we also verify that C/libbpf still builds and loads the buggy code.
% 对于三个 prior work 案例，我们还验证 C/libbpf 仍能构建并加载错误代码。
These cases are an XDP program typed with \texttt{\_\_sk\_buff}, a 16-byte update into an 8-byte map value, and a map lookup reinterpreted as an unrelated \texttt{u64}.
% 案例为：以 \texttt{\_\_sk\_buff} 为类型的 XDP 程序、向 8 字节映射值写入 16 字节更新，以及对查找结果做无关 \texttt{u64} 重解释。
C buggy objects build with clang, pass verifier load, and reproduce the bug under runtime checks: map cases use \code{BPF_PROG_TEST_RUN} to observe truncation or reinterpretation, and the context case attaches on a veth to observe field misalignment under traffic.
% C 错误对象用 clang 构建、通过验证器加载，并在运行时检查下复现 bug：映射案例用 \code{BPF_PROG_TEST_RUN} 观察截断或重解释，上下文案例在 veth 上附加以观察流量下的字段错位。
% 上下文案例在新建 veth 上附加，驱动跨命名空间流量，并检查错名的 \texttt{protocol}/\texttt{queue\_mapping} 读数等于对应偏移上的 XDP 布局字段（\texttt{rx\_queue\_index}/\texttt{ingress\_ifindex}），且错名 protocol 值不是以太网类型 \texttt{0x0800}。

\begin{table}[t]
\caption{Compile-time rejection versus continued load for the three cases
above. \tool{} rejects all three at compile time; C/libbpf builds and
loads them, and only a runtime check observes the bug.}
% 上述三个案例的编译期拒绝与持续加载对比。\tool{} 均在编译期拒绝；C/libbpf 构建并加载它们，仅运行时检查观测到 bug。
\label{tab:q1-compile}
\small
\begin{tabular}{@{}p{0.27\columnwidth}p{0.26\columnwidth}p{0.33\columnwidth}@{}}
\toprule
Case & \tool{} & C/libbpf \\
\midrule
XDP program typed with \code{*__sk_buff} & compile reject: wrong context signature & loads; misnamed fields read XDP layout on veth \\
\addlinespace
16-byte update into 8-byte map value & compile reject: map value type mismatch & loads; value truncated under \code{BPF_PROG_TEST_RUN} \\
\addlinespace
Map lookup read as unrelated \code{u64} & compile reject: declaration type mismatch & loads; native-endian reinterpretation under \code{BPF_PROG_TEST_RUN} \\
\bottomrule
\end{tabular}
\end{table}

Table~\ref{tab:q1-compile} shows the same pattern on all three cases. \tool{} rejects at compile time, while C/libbpf loads and requires a runtime check to observe the bug.
% 表~\ref{tab:q1-compile} 在全部三个案例上显示同一模式：\tool{} 编译期拒绝，C/libbpf 仍加载并用运行时检查暴露错误。
On the context case, C/libbpf accepts the wrong-typed object. Under veth traffic, the misnamed fields return XDP \texttt{ingress\_ifindex}/\texttt{rx\_queue\_index} values rather than TC protocol/queue values.
% 在上下文案例中，C/libbpf 接受类型错误的对象，并在 veth 流量下使错名字段观测到 XDP 的 \texttt{ingress\_ifindex}/\texttt{rx\_queue\_index} 而非 TC 的 protocol/queue 语义。
Map cases reproduce truncation and native-endian reinterpretation under \code{BPF_PROG_TEST_RUN}.
% 映射案例在 \code{BPF_PROG_TEST_RUN} 下复现截断与本地端序重解释。

\FloatBarrier
\subsection{Q2: How Much Smaller Is \tool{} Code than C/libbpf?}
% Q2：\tool{} 代码比 C/libbpf 小多少？

\begin{figure}[t]
\centering
\includegraphics[width=\columnwidth]{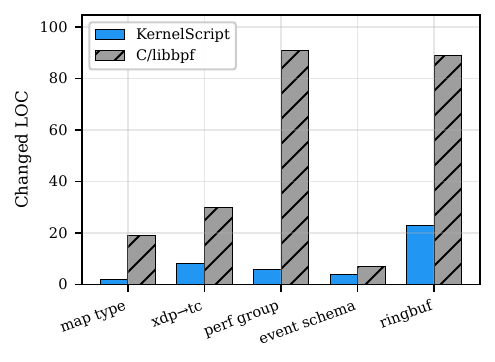}
\caption{Changed LOC for the same five changes.
\tool{} stays in a single typed source, while C/libbpf requires multi-file
coordination (median 6 vs.\ 30 LOC, and \tool{} needs no manual
kernel/userspace/header sync on any change).}
% 相同五个变更的变更 LOC。\tool{} 留在单一类型化源内，而 C/libbpf 需要多文件协调（中位 6 对 30 LOC，且 \tool{} 任一变更均无需手动内核/用户空间/头文件同步）。
\label{fig:q2-change}
\end{figure}

Across 11 workloads with hand-written C/libbpf baselines, \tool{} sources total 203 lines versus 1105 for C/libbpf (5.4$\times$ smaller).\footnote{SLOC counts exclude blank lines, comments, and generated headers.}
% 在 11 个带手写 C/libbpf 基线的工作负载上，\tool{} 源共 203 行，而 C/libbpf 为 1105 行（小 5.4 倍）。SLOC 计数排除空行、注释和生成的头文件。
Cross-boundary changes are also smaller: we construct five changes (map \code{array}$\rightarrow$\code{percpu\_array}, program \code{@xdp}$\rightarrow$\code{@tc}, perf\_event standalone$\rightarrow$grouped, shared ring-buffer event struct, adding ring-buffer reporting) and compare diffs.
% 为衡量该成本，我们构造相同的五个变更（映射 \code{array}$\rightarrow$\code{percpu_array}、程序 \code{@xdp}$\rightarrow$\code{@tc}、perf\_event 独立$\rightarrow$分组、共享 ring-buffer 事件结构体，以及添加带用户空间处理的 ring-buffer 报告），并比较 \tool{} 与手写 C/libbpf 的 diff。
Figure~\ref{fig:q2-change} plots changed LOC per change.
% 图~\ref{fig:q2-change} 绘制每次变更的 LOC。
We also record files touched, distinct edit sites, and which sides need manual updates, whether kernel, userspace, or generated headers and BPF section names.
% 我们还记录触及的文件数、不同修改点，以及哪些侧需要手动更新（内核、用户空间，或生成头文件与 BPF section 名）。

None of the \tool{} changes require manual cross-boundary synchronization, whereas hand-written C/libbpf requires userspace synchronization in 5/5 cases, kernel-side synchronization in 4/5, and generated-header or BPF section-name updates in 3/5.
% \tool{} 的变更都不需要手动跨边界同步，而手写 C/libbpf 在 5/5 个案例中需要用户空间同步，在 4/5 个案例中需要内核侧同步，在 3/5 个案例中需要生成头文件或 BPF section 名更新。
\tool{} touches 1 file, 2 edit sites, and 6 lines (median), versus 2, 7, and 30 for C/libbpf.
% \tool{} 触及 1 个文件、2 个修改点和 6 行（中位数），而 C/libbpf 为 2、7 和 30。
As cross-boundary structure accumulates (XDP that only forwards packets → counter map → ring-buffer reporting), \tool{} grows more slowly: +2 lines for the counter map versus +11 in C/libbpf, and +12 for ring-buffer reporting versus +202.
% 随着跨边界结构累积（仅转发数据包的 XDP → 计数映射 → ring-buffer 报告），\tool{} 增长更慢：计数映射 +2 行相对 C/libbpf 的 +11，ring-buffer 报告 +12 相对 +202。
By the reporting stage, C/libbpf reaches 9$\times$ the \tool{} source size.
% 到报告阶段，C/libbpf 达到 \tool{} 源大小的 9 倍。

\subsection{Q3: Is \tool{} Compatible with clang, bpftool, and libbpf?}
% Q3：\tool{} 是否与 clang、bpftool 和 libbpf 兼容？
\label{sec:eval:q3}

Of 43 repository examples spanning XDP, TC, kprobe, tracepoint, perf\_event, maps, ring buffers, tail calls, kfuncs, and struct\_ops, 41 build unmodified through the generated Makefile and the remaining two struct\_ops examples after a one-line fix for a libbpf version skew: bpftool 7.7's generated userspace header assigns a map \code{link} field that the installed libbpf 1.3.0 does not define (libbpf 1.6 does), so the incompatibility lies in bpftool's output rather than in code \tool{} emits.
% 在涵盖 XDP、TC、kprobe、tracepoint、perf\_event、映射、环形缓冲区、尾调用、kfunc 和 struct\_ops 的 43 个仓库示例中，41 个经生成 Makefile 直接构建成功，其余两个 struct\_ops 示例在针对 libbpf 版本错配做一行修正后构建成功：bpftool 7.7 生成的用户空间头文件赋值了一个映射 \code{link} 字段，而已安装的 libbpf 1.3.0 未定义该字段（libbpf 1.6 定义了），因此该不兼容位于 bpftool 的输出而非 \tool{} 发出的代码。
Both then load, attach, and detach correctly on a real kernel.
% 两者随后在真实内核上均正确加载、附加与分离。
On a real kernel, 27 single-program XDP objects attach and detach in isolated netns/veth pairs.
% 在真实内核上，27 个单程序 XDP 对象在隔离 netns/veth 对中成功附加与分离。
The same workloads for XDP/TC traffic, ring-buffer events, perf\_event counters, and selected struct\_ops paths pass the same behavioral checks.
% XDP/TC 流量、ring-buffer 事件、perf\_event 计数器与选定 struct\_ops 路径上的相同工作负载在相同行为检查下一致。
Under \code{BPF_PROG_TEST_RUN}, a minimal generated XDP pass runs at 5\,ns per invocation against 6\,ns for its hand-written baseline.
% 在 \code{BPF_PROG_TEST_RUN} 下，最小生成的 XDP pass 每次调用 5\,ns，手写基线为 6\,ns。
Across ten veth/iperf3 trials, generated objects sustain 17.25\,Gbps against 17.34 hand-written on a counter workload and 17.35 against 17.81 on pass, within run-to-run variance.
% 在十次 veth/iperf3 试验中，生成对象在计数工作负载上维持 17.25\,Gbps，手写为 17.34；pass 上为 17.35 对 17.81，均在运行间波动范围内。

\section{Related Work}
% 相关工作

Existing work on eBPF development falls into three categories.
% 现有 eBPF 开发相关工作分为三类。
Specialized DSLs target particular use cases, including bpftrace for tracing~\cite{bpftrace}, BPFBox and ActPlane for policy enforcement~\cite{bpfbox,actplane}, BMC for caching~\cite{bmc}, P4c-eBPF for networking~\cite{p4c-ebpf}, and Yaksha-Prashna for bytecode analysis~\cite{yaksha}.
% 专用 DSL 针对特定用例，包括用于追踪的 bpftrace、用于策略执行的 BPFBox 和 ActPlane、用于缓存的 BMC、用于网络的 P4c-eBPF，以及用于字节码分析的 Yaksha-Prashna。
These DSLs primarily target kernel-side code and do not type the full cross-boundary structure that \tool{} addresses.
% 这些 DSL 主要针对内核侧代码，不对 \tool{} 所处理的完整跨边界结构进行类型化。
General eBPF development tools include the standard eBPF development workflow~\cite{libbpf}, Aya for Rust~\cite{aya}, Rex for safe Rust kernel extensions without a verifier~\cite{rex}, Wasm-bpf for WebAssembly deployment~\cite{wasmbpf}, and eunomia-bpf for dynamic loading~\cite{eunomia-bpf}.
% 通用 eBPF 开发工具包括标准 eBPF 开发工作流~\cite{libbpf}、用于 Rust 的 Aya~\cite{aya}、不使用验证器的安全 Rust 内核扩展 Rex~\cite{rex}、用于 WebAssembly 部署的 Wasm-bpf~\cite{wasmbpf}，以及用于动态加载的 eunomia-bpf~\cite{eunomia-bpf}。
These reduce boilerplate but do not type execution domains or enforce context/return-type constraints at compile time.
% 这些减少样板但不对执行域进行类型化，也不在编译期强制上下文/返回类型约束。
Verified approaches take a different direction: BeePL provides correct-by-compilation kernel extensions~\cite{beepl}, and Jitk and Serval verify JIT correctness~\cite{jitk,serval}.
% BeePL 提供编译即正确的内核扩展~\cite{beepl}，Jitk 和 Serval 验证 JIT 正确性~\cite{jitk,serval}。
\tool{} differs by generating both kernel and userspace code from one source and typing cross-boundary relationships. It aims for practical integration with the existing toolchain rather than formal verification.
% \tool{} 不同：从一个源生成内核与用户空间代码并对跨边界关系进行类型化，目标是与现有工具链的实际集成，而非形式化验证。

\section{Conclusion}
% 结论

\tool{} types the cross-boundary structure of an eBPF application, covering program lifecycle, shared maps, and execution domains.
% \tool{} 对 eBPF 应用程序的跨边界结构进行类型化，涵盖程序生命周期、共享映射和执行域。
This replaces conventions and late verifier or attach-time failures with compile-time checks.
% 这用编译期检查取代约定以及延迟的验证器或附加时故障。
Typed checks reject wrong context types and map type bugs before code generation. C/libbpf still builds and loads these bugs, exhibiting the corresponding runtime misinterpretation.
% 类型检查在代码生成前拒绝错误上下文类型与映射类型 bug，而 C/libbpf 仍构建、加载并表现出相应运行时误解释。
A unified source keeps the same cross-boundary changes smaller, and generated objects build and run through clang, bpftool, and libbpf on a real kernel.
% 统一源使同一跨边界修改更小，生成对象在真实内核上经 clang、bpftool 和 libbpf 构建并运行。

\newpage
\bibliographystyle{ACM-Reference-Format}
\bibliography{references}

\end{document}